\documentclass{iopart}
\usepackage{graphicx}
\input{epsf}
\usepackage{amssymb}
\begin{document}

\title{Quantum Gibbs distribution from dynamical thermalization
in classical nonlinear lattices} 
\author{Leonardo Ermann$^1$ and Dima L. Shepelyansky$^2$}
\address{$^1$ Departamento de F\'\i sica Te\'orica, GIyA, Comisi\'on Nacional 
de Energ\'ia At\'omica, Buenos Aires, Argentina}
\address{$^2$Laboratoire de Physique Th\'eorique du CNRS, IRSAMC, 
Universit\'e de Toulouse, UPS, 31062 Toulouse, France}

\begin{abstract}
We study numerically time evolution in classical lattices with 
weak or moderate nonlinearity which leads 
to interactions between linear modes. Our results show that 
in a certain strength range 
a moderate nonlinearity generates a
dynamical thermalization process which drives the system
to the quantum Gibbs distribution of probabilities,
or average oscillation amplitudes. 
The effective dynamical temperature of the lattice
varies from large positive to large negative values
depending on energy of initially excited modes.
This quantum Gibbs distribution
is drastically different from 
usually expected  energy equipartition over linear modes
corresponding to a regime of
classical thermalization.
Possible experimental observations
of this dynamical thermalization are discussed
for cold atoms in optical lattices,
nonlinear photonic lattices and optical fiber arrays.  

\end{abstract}


\vskip 0.3cm
Dated: July 22, 2013

\vskip 0.3cm
\submitto{\NJP}

\maketitle
\section{Introduction}

The problem of thermal distribution for photons
led to the invention of the Planck constant
and Planck law \cite{planck}. Further 
development of quantum mechanics 
generalized the Gibbs thermal distribution \cite{gibbs}
to the quantum case leading to the quantum Gibbs
distribution in a quantum system with
discrete energy levels (see e.g. \cite{landau,mayer}). 
Thus the problem of thermalization 
was always fascinating the scientists
starting from the famous dispute 
between Boltzmann and Loschmidt
on time reversibility and
statistical description (see e.g.
\cite{mayer}).

The thermalization in a given system
is based on the ergodicity 
of motion which can be produced by noise from a 
heat bath or by  internal dynamical
chaos. The mathematical and physical
foundations of dynamical chaos
are now well established and
are described in \cite{arnold,sinai,chirikov,lichtenberg}.
The first numerical investigations
of onset of ergodicity and dynamical thermalization
in a nonlinear lattice of coupled oscillators
had been performed for the Fermi-Pasta-Ulam
problem \cite{fpu1,fpu2,fpu3,fpu4} with an expectation
to find energy equipartition over 
linear oscillator modes. 
Surprisingly, for a typical
set of parameters
the equipartition
was absent, even if in certain
cases signs of non-periodic 
behaviour were visible.
The absence of ergodicity
stimulated a great interest to the Fermi-Pasta-Ulam problem
even if later it became clear that this model is rather close
to the integrable Toda lattice and, hence,
it does not belong to a class of generic models
(see discussions in \cite{lichtenberg,fpu3,fpu4}). 

Another approach to investigation of
onset of ergodicity over linear oscillator modes
in nonlinear lattices
had been proposed in \cite{dls1993}
by analyzing the effects of nonlinearity
on the Anderson localization \cite{anderson}
in systems with disorder or systems of quantum chaos.
It was found that below a certain critical nonlinearity
a spreading over modes is suppressed or is exponentially slow
while at moderate nonlinearity
a sub-diffusive spreading continues up to times
being by millions time larger
than a typical time scale of oscillations.
This result has been confirmed and
significantly extended by further investigations
\cite{molina,danse,flach2009prl,flach2009kg,garciapre,flach2010pre,
flach2010epl,aubry2010,marioepl2010,fishmanpre2011,marionjp2013,flach2013},
however the full understanding of the problem is 
still lacking. Thus the results \cite{fishmanpre2011}
indicate that at large times the spearing continues
along certain chaotic but non-ergodic layers.
The mathematical studies \cite{wang,bourgain,fishman2012}
demonstrate all the complexity of this problem
where pure-point spectrum of linear system
generates intricate resonances induced by nonlinearity. 
The interest to the problem is also supported by
experiments with disordered nonlinear photonic lattices
\cite{segev,silberberg} and 
Bose-Einstein condensates of cold atoms 
placed in a disordered optical lattice \cite{inguscio}.

Recently it was  argued that in the discrete Anderson
nonlinear Sch\"odinger equation (DANSE) a process of
dynamical thermalization takes place leading to
a statistical equilibrium in a finite disordered
lattice at a moderate nonlinearity 
\cite{dynthermo}. It was shown numerically that
the Gibbs energy distribution
takes place over linear eigenmodes.
This work generated a certain interest to the
process of dynamical thermalization in 
weakly nonlinear lattices \cite{kottos}.
It was also pointed out that such a thermalization
is necessary for emergence of Kolmogorov
turbulence in finite size systems \cite{dlskolm}.
 
Here we extend the studies of dynamical thermalization
in disordered lattices with weak or moderate nonlinearity.
We especially stress the situation when the energies
of linear modes grow linearly with index of linear modes
corresponding to a static Stark field or finite
density of levels in a unit energy (frequency) interval.
Such a case is typical for the Kolmogorov (or weak wave) turbulence
in finite systems \cite{zakharov,nazarenko}.
As an example of such a system we can name 
the nonlinear Schr\"odinger equation
in the Sinai billiard (or any other chaotic billiard)
as discussed in \cite{dlskolm}.
It is also important to note that the DANSE with a static
field is also characterized by a subdiffusive
spreading \cite{garciastark}.

In this work we extend the research line 
of dynamical thermalization in nonlinear disordered lattices
investigating a large number of models.
Surprisingly, our results show that
in  lattices with weak or moderate nonlinearity
there is emergence of a quantum Gibbs distribution
over energies of linear eigenmodes.
In some sense the weak nonlinearity
acts as a dynamical thermostat
creating a quantum Gibbs distribution.
We discuss the conditions under which 
such a quantum Gibbs replaces a usually
expected energy equipartition over
linear modes predicted by 
the classical thermalization theory
\cite{landau,mayer,arnold,sinai,chirikov,lichtenberg}.

The paper is constructed as follows: in Section 2 we describe 
all nonlinear lattice models investigated in this work,
in Section 3 we introduce the quantum Gibbs anzats,
results for 1d models $M1, M2$ and 2d  models $M3, M4$ 
are presented in Sections 4 and 5,
the results for the Klein-Gordon lattice are given in Section 6,
the discussion of the results is presented in Section 7.  

\section{Description of nonlinear lattice models }

To investigate the phenomenon
of emergence of a quantum Gibbs distribution
we study several models of linear 
lattices with disorder and additional weak
or moderate nonlinear terms. These models
represent one-dimensional (1d) and
two-dimensional lattices (2d)
which in absence of nonlinearity can be reduced
to the Anderson model 
of non-interacting electrons (see e.g. \cite{mirlin}) on
a disordered lattice in 1d and 2d respectively. 

The main DANSE model \cite{danse} is described by the equation:
\begin{equation}
 i \hbar{{\partial {\psi}_{n}} \over {\partial {t}}}
=E_{n}{\psi}_{n} 
+{\beta}{\mid{\psi_{n}}\mid}^2 \psi_{n}
 +V ({\psi_{n+1}}+ {\psi_{n-1})} \; .
\label{eq1} 
\end{equation} 
In the following we use dimensionless units with
$\hbar=V=1$, the Boltzmann constant is taken to be unity
so that we have all dimensionless variables. 
In total we consider the lattice with $N$ 
sites and periodic boundary conditions.
For $E_n=0$ and long wave limit the
system is reduced 
to the nonlinear Schr\"odinger equation
which is also known in the field of cold atoms as the Gross-Pitaevskii
equation \cite{nse}. At $\beta=0$ and random values of $E_n$
distributed in the interval $-W/2 \leq E_n \leq W/2$
the system (\ref{eq1}) represents the 1d Anderson model
with the localization length $\ell \approx 96/W^2$ \cite{mirlin}.
For this distribution of $E_n$ and nonzero $\beta$ 
the equation (\ref{eq1}), named as the DANSE model,
was discussed and 
investigated in \cite{dls1993,molina,danse,flach2009prl,flach2009kg}
and other papers.

The Hamiltonian of DANSE has the form
\begin{equation}
H=\sum_{n}E_n|\psi_n|^2+\psi_{n-1}\psi_n^*+
\psi_{n-1}^*\psi_n+\frac{\beta}{2}|\psi_n|^4\; ,
\label{eq2}
\end{equation}
with $\psi_n$ and $\psi_n^*$ being the conjugated variables.
The energy and the probability norm $\sum_n |\psi_n|^2=1$
are exact integrals of motion. The Hamiltonian (\ref{eq2})
can be rewritten in the basis of linear eigenmodes
 $\varphi_{nm}$ related to $\psi_n=\sum C_m\varphi_{nm}$.
In the eigenmode representation the Hamiltonian
is
\begin{equation}
H=\sum_{m=1}^N \epsilon_m |C_m|^2+
\beta\sum_{m_1m_2m_3m}V_{m_1m_2m_3m}C_{m_1}C_{m_2}C^*_{m_3}C^*_{m},
\label{eq3}
\end{equation}
with $\sum_m |C_m|^2=1$, and $V_{{m}{m'}{m_1}{m_1'}} \sim \ell^{-3/2}$ 
being the transition matrix elements \cite{dls1993}
(the dependence on $\ell$ is given assuming random matrix 
estimate for eigenstates overlap). 
From this representation it is especially clear that the
spreading takes place only due to the nonlinear $\beta$ coupling.
 
In 1d we consider the extensions of the DANSE model
given by the following replacements in Eq.~(\ref{eq1}):
\begin{equation}
E_n \rightarrow E_n+ f|n-n_0| \; , \; (M1) \; .
\label{eq4}
\end{equation}
Here $E_n$ have the same random distribution as in DANSE,
$n_0=(N+1)/2$ marks the center of the lattice
and the periodic conditions link sites $N$ and $1$.
This is the model $M1$ with the static Stark field $f$
which models the constant density of states in energy
as it is the case in the quantum Sinai billiard \cite{dlskolm}.

We also study the model $M2$ which is obtained from $M1$
by the following replacement of the nonlinear term:
\begin{equation}
\beta \rightarrow \beta |n-n_0| \; , \; (M2) \; .
\label{eq5}
\end{equation}
In this model $M2$ the nonlinear term
grows with the level number
that often happens for nonlinear wave interactions
in wave turbulence (see e.g. \cite{zakharov,nazarenko,dlsfpu}).

We also analyze the 2d DANSE lattice
studied in \cite{garciapre}:
\begin{eqnarray}
\nonumber
 i {{\partial {\psi}_{n_x n_y}} \over {\partial {t}}}
=E_{n_x n_y}{\psi}_{n_x n_y} 
+{\beta}{\mid{\psi_{n_x n_y}}\mid}^2 \psi_{n_x n_y} \\
 + ({\psi_{n_x+1 n_y}}+ {\psi_{n_x-1 n_y}}+ {\psi_{n_x n_y+1}}+  {\psi_{n_x n_y-1}}) \; .
\label{eq6} 
\end{eqnarray}
Periodic boundary conditions are used for $N \times N$ square lattice
with $-N/2 \leq n_x , n_y \leq N/2$.
However, here we use the extended version of this model
assuming that
\begin{equation}
E_{n_x n_y} = \delta E_{n_x n_y}+ f (n_x^2+n_y^2)\; , \;  
-W/2 \leq \delta  E_{n_x n_y} \leq W/2\; , \; (M3) \; .
\label{eq7}
\end{equation}
This is the model M3 with random values of 
energies $\delta  E_{n_x n_y}$ in a given interval.

In addition we study the model $M4$ obtained from the model $M3$
by the replacement
\begin{equation}
\beta \rightarrow \beta (n_x^2+n_y^2) \; , \; (M4) \; .
\label{eq8}
\end{equation}
This is the 2d analog of model $M2$.

Since the term $(n_x^2+n_y^2)$ grows linearly
with index $k = |n_x|+|n_y|$ we can consider the model
$M3$ as the model for the nonlinear Schr\"odinger
equation in the Sinai billiard (see Eq.(6) at $F=0$ in \cite{dlskolm}).
Indeed,  in a Sinai billiard the energy levels 
are randomly and homogeneously distributed over the energy
axis, as it is the case in model $M3$ at $f>0$,
 and also the nonlinear term has a similar form coupling the
linear modes. The advantage of $M3$ model is that it is 
significantly easier for numerical simulations compared to the
case of Sinai billiard. The model $M4$ has a stronger nonlinear
interactions at high wave vectors that is typical for 
the weak wave turbulence \cite{zakharov,nazarenko}.

We note that 2d models $M3$, $M4$ also can be written in the form 
 (\ref{eq3}) with more complex matrix elements induced by 
the nonlinear coupling on 2d lattice.

The above models $M1, M2, M3, M4$ have two integrals of motion
being energy and the wavefunction norm.
The latter is generally absent in nonlinear lattices.
For this reason we consider the Klein-Gordon lattice
(KG model) described by the Hamiltonian:
\begin{equation}
H=\sum_{l}[(p^2_l+{\tilde \epsilon_l} u_l^2)/2+\beta u_l^2/4 +
(u_{l+1}-u_l)^2/(2W)] ,
\label{eq9}
\end{equation}
where $\tilde \epsilon_l$ are taken as random in
the interval $[1/2,3/2] $ (see e.g. \cite{flach2009kg}).
This KG model was studied in \cite{flach2009kg}
and it was shown that it has the same type 
of subdiffusive spreading as DANSE.
We keep the same notations as in \cite{flach2009kg}
(see Eq.(6) there) but we introduce
the nonlinear coefficient $\beta$
(it is taken at $\beta=1$ in \cite{flach2009kg})
and we add a static field $f$
replacing $\tilde \epsilon_l \rightarrow \tilde \epsilon_l +f |l-l_0|$
keeping the random distribution in the same
interval (in \cite{flach2009kg} $f=0$).
We use $l_0=(N+1)/2$ and periodic conditions
linking sites $l=1$ and $N$.
As shown in \cite{flach2009kg}, 
the linear part of the Hamiltonian at $f=0, \beta=1$
can be reduced to the 1d Anderson model.

The time evolution of models $M1, M2, M3, M4$ 
was integrated numerically using
 the symplectic integration scheme as described in \cite{garciapre}; 
the KG model was integrated by
$SABA_2C$ method described in \cite{flach2009kg}.
The time average is done over the time interval $\delta t$ in
a vicinity of time $t$.
The integration time step was fixed at 
$\delta t=0.05$ for all models
but we checked that its decrease by a factor $ 10 $
did not affect the results of numerical simulations.

\section{Quantum Gibbs anzats}

For the DANSE and $M1, M2, M3, M4$ models we 
make a quantum Gibbs conjecture that 
the nonlinear terms act like some kind of 
dynamical thermostat
which creates the quantum Gibbs distribution over
quantum states with linear mode  eigenenergies $\epsilon_m$.
Then according to the standard relations 
of statistical mechanics \cite{landau,mayer}  
we find the probabilities $\rho_m=|C_m|^2$
and the statistical sum $Z$ of the system:
\begin{equation}
\rho_m=Z^{-1} \exp(-\epsilon_m/T) \; , \; Z=\sum_m \exp(-\epsilon_m/T) \; .
\label{eq10}
\end{equation}
Here, $T$ is a certain temperature of our isolated
system which depends on the initial energy
given to the system. As usually for any quantum system with
energy levels $\epsilon_m$ we have
the total probability $\sum_ m \rho_m=1$
and total energy $E=\sum \rho_m \epsilon_m$
(here we neglect a small nonlinear term correction to energy).
The norm conservation can also taken into account 
using the standard approach of statistical mechanics
with the chemical potential and conservation of number of particles
(or norm) \cite{landau,mayer} that is equivalent to the
normalization used in (\ref{eq10}). We note that
possibilities of thermalization has been discussed in
nonlinear chains starting from the FPU problem \cite{fpu1,fpu2,fpu3,fpu4}
and continuing even for nonlinear breathers 
\cite{breather1,breather2,rumpf}. However, here we consider the case of weak
or moderate nonlinearity when the nonlinear terms
are relatively small comparing to linear quadratic terms.
In this case the classical system  is expected to reach
energy equipartition over linear modes \cite{landau,mayer,tolman,henry}.

The entropy of the system can be expressed via 
the average probability $\rho_m$ on level $m$
via the usual formula: 
\begin{equation}
S = -\sum_m \rho_m \ln \rho_m \; , \qquad \rho_m=\overline{|C_m|^2} \; ,
\label{eq11}
\end{equation}
where overline means time averaging.

\begin{figure} 
\begin{indented}\item[]
\begin{center}
\includegraphics[width=0.80\textwidth]{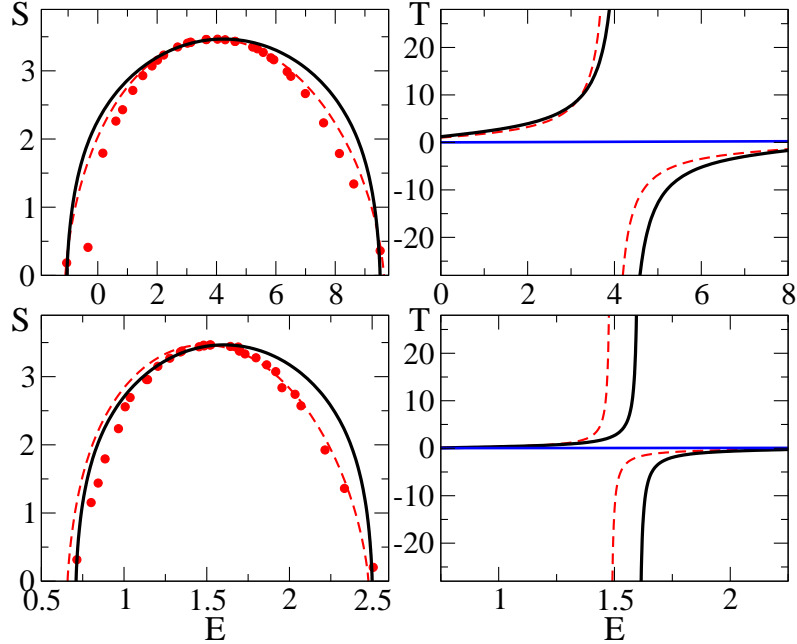}
\end{center}
\vglue -0.2cm
\caption{Top panels: (left) the dependence of
entropy $S$ on energy $E$ 
from the quantum Gibbs anzats (\ref{eq10}) with 
$\epsilon_m$ taken from a given disorder realisation in $M1$
(red dashed curve), the numerical data
from time evolution of $M1$ are shown by red points
at $t=10^7$ with $\delta t=10^6$;
the full black curve
shows the dependence from the Gibbs anzats (\ref{eq10})
for equidistant levels $\epsilon_m={\tilde f} m/2$
with ${\tilde f} =2(\epsilon_{max}-\epsilon_{min})/N \approx 1.32 f  $
(where $\epsilon_{max}=9.51$ and $\epsilon_{min}=-1.05$)
for a given disorder realization (see text);
here $f=0.5$, $\beta=2$, $W=2$, $N=32$;
(right) temperature dependence $T(E)$ 
from the quantum Gibbs anzats (\ref{eq10})
with the same cases as in left panel, blue curve shows
the classical equipartition dependence 
$T=E/N$, numerical data are not shown here.
Bottom panels: 
same as in the top panels but for the $KG$ model
at $f=0.125$, $\beta=1$, $W=2$, $N=32$, 
$\epsilon_{max}=2.51$ and $\epsilon_{min}=0.71$,  
numerical evolution is followed up to $t=10^8$ with $\delta t=10^6$.
}
\label{fig1}
\end{indented}
\end{figure} 

The entropy $S$, energy $E$ and temperature $T$ are related to each other 
via the standard thermodynamics expressions \cite{landau}:
\begin{equation}
E= T^2 \partial \ln Z/\partial T\;, \;\; S= E/T+\ln Z \;, \;\; 
\partial S/ \partial E =1/T \; .
\label{eq12}
\end{equation}
This value of entropy yields the maximal possible 
equipartition for a given initial energy.
In an implicit way, a value of energy $E$  
determines the temperature $T$ of the system
and its entropy, or by varying temperature 
$T$ in the range $(-\infty, +\infty)$
we obtain the variation $E(T)$, $S(T)$
and implicitly the curve $S(E)$.
The advantage of variables $E, S$
is based on the fact that they both
are extensive variables \cite{landau,mayer}
and thus they are self averaging
and hence in numerical simulations
they have significantly smaller fluctuations
comparing e.g. to temperature $T$.
It is important to note that the above
quantum Gibbs relations can be also obtained from the
condition that the entropy $S$ takes
the maximal value at variation of 
probabilities $\rho_m$.

In fact the quantum Gibbs anzats was introduced 
in \cite{dynthermo} for the DANSE and it was shown
that it works at moderate nonlinearity $\beta$
and not very strong disorder $W$ 
(see also discussions in \cite{rumpf}).
However, in \cite{dynthermo} the {\it striking
paradox of quantum Gibbs anzats} was not 
pointed out directly.
Indeed, the nonlinear classical lattice
is expected to have energy equipartition over
linear modes that is in a drastic contrast with the
quantum Gibbs distribution described above.

\begin{figure} 
\begin{indented}\item[]
\begin{center}
\includegraphics[width=0.80\textwidth]{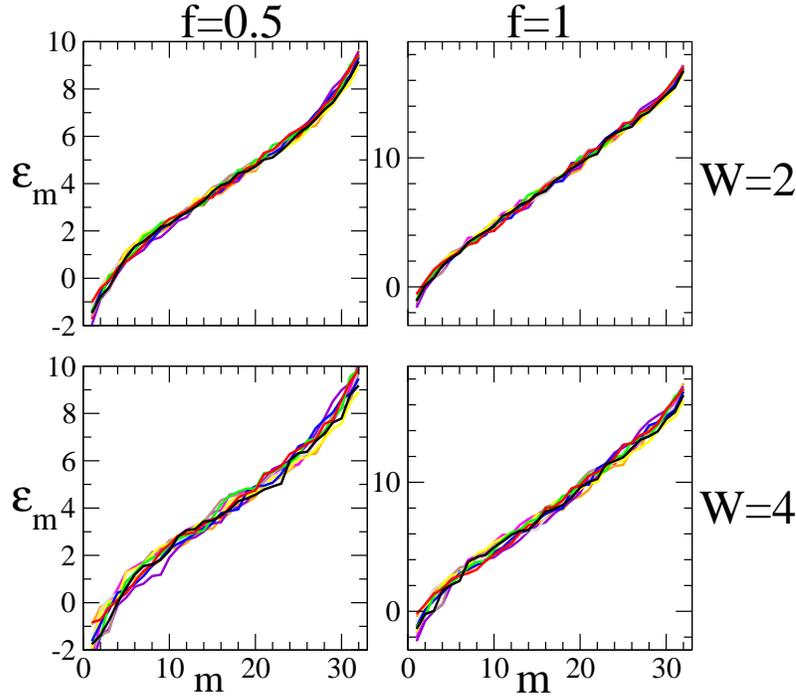}
\end{center}
\vglue -0.2cm
\caption{Dependence of eigenenenergies $\epsilon_m$ of linear eigenmodes
on mode index $m$ in $M1$
for parameters of Fig.~\ref{fig1} at $f=0.5; 1$
(ten disorder realisations are shown by different color curves).
}
\label{fig2}
\end{indented}
\end{figure}

The examples of dependence $S(E)$ and $T(E)$
produced by the quantum Gibbs anzats for the models $M1$ and $KG$
are shown in Fig.~\ref{fig1}. We use one disorder
realisation with eigenvalues $\epsilon_m$
for $M1$ and $\epsilon_m=\omega_m^2/2$ for $KG$
(more details on KG model are given in Section 6).
To compare the numerical data obtained from time evolution
with the Gibbs anzats we use the exact eigenenergies
$\epsilon_m$ obtained from exact matrix diagonalization of the linear
problem at a given disorder realisation.
Examples of dependence of $\epsilon_m$ on index $m$
are shown in Fig.~\ref{fig2} for the model $M1$.
We also can use the average dependence 
$\epsilon_m \approx { \tilde f} m/2$ ($1 \leq m \leq N$)
with ${\tilde f} = 2(\epsilon_{max}-\epsilon_{min})/N$
which in an approximate manner takes
into account the disorder fluctuations
with the maximal $\epsilon_{max}$ and minimal
values of $\epsilon_{min}$ linear eigenenergies. 
This approach of an effective  
average density ${\tilde f}$ 
gives a good description of numerical data (see Fig.~\ref{fig1}).
It gives a slight shift of the maximum of $S(E)$ curve
which is more sensitive to a disorder and is not
of principal importance.
We return to the discussion of $KG$ model in Section 6.

In contrast to the quantum Gibbs distribution the classical
thermodynamics implies the energy  equipartition over all
modes \cite{landau,mayer} that gives:
\begin{equation}
T=(E-E_{min})/N \; , \;\; S =N \ln (E-E_{min})+ C_0 \; ,
\label{eq13}
\end{equation}
where $E_{min}$ is a certain minimal energy of the system
and $C_0$ is a numerical constant.
The results of Fig.~\ref{fig1} show the drastic difference between
the predictions of quantum and classical thermodynamics.

The dependence $S(E)$ has one maximum and according to the standard 
thermodynamics relations (\ref{eq12}) the system has a negative
temperature $T<0$ at the right branch of $S(E)$ curve. 
It is known that such situations can appear in quantum systems
with energies located in a finite band width \cite{landau,mayer}.
We note that recent experiments 
with cold atoms in optical lattices \cite{braun}
allowed to realise finite quantum systems
at negative temperatures. 

We should stress that  the quantum Gibbs distribution we find
has close similarities with the thermal quantum distribution
in real quantum systems however it appears as a result of dynamical
thermalization in weakly nonlinear classical coupled
oscillators without any second quantization.
This Gibbs distribution
results from dynamical thermalization
and entropy maximization over linear modes
without real quantum Plank constant entering in the game.
In this respect our physical interpretation is very 
different from the one developed
in \cite{galgani} where the authors discussed
appearance of the real quantum Planck constant in the
thermal equilibrium of classical nonlinear lattices.
In our consideration we have an effective Planck
constant which may be effectively introduced in
a system of weakly coupled nonlinear oscillators
(e.g. as a typical frequency difference between frequencies
for DANSE or KG models).

Below we present the numerical results on the 
detailed verification 
of the quantum Gibbs anzats for various lattice models.

\section{Results for 1d lattice models }

\begin{figure} 
\begin{indented}\item[]
\begin{center}
\includegraphics[width=0.80\textwidth]{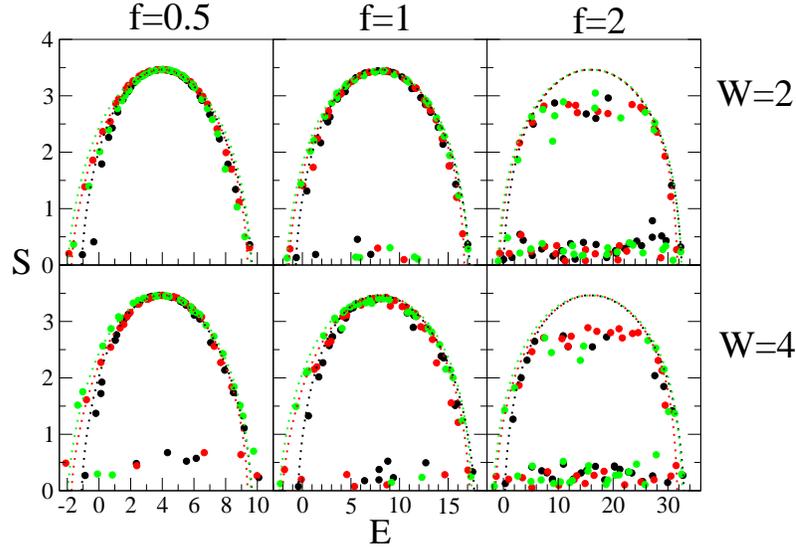}
\end{center}
\vglue -0.2cm
\caption{Dependence of entropy $S$ on energy $E$ 
for  initial excited eigenmodes of the linear problem in 1d model $M1$.
The value of entropy $S$ is obtained
by time averaging over a time interval $\delta t=10^6$,
 at time $t=10^7$, the energy $E$
is taken as the total energy of the lattice
when the linear eigenmode $m$ is
excited at time $t=0$ ($1\leq m \leq N$).
The numerical data are shown by points for 
3 disorder realisations  (black, green, red), 
the corresponding theoretical 
quantum Gibbs distributions, 
with the exact linear eigeneinergies $\epsilon_m$
for a give disorder realisation,
are shown by dotted curves of the same color.
The values of disorder strength $W$ 
and Stark field $f$ are given in the figure;
here $\beta=2$, $N=32$.
}
\label{fig3}
\end{indented}
\end{figure}

\begin{figure} 
\begin{indented}\item[]
\begin{center}
\includegraphics[width=0.80\textwidth]{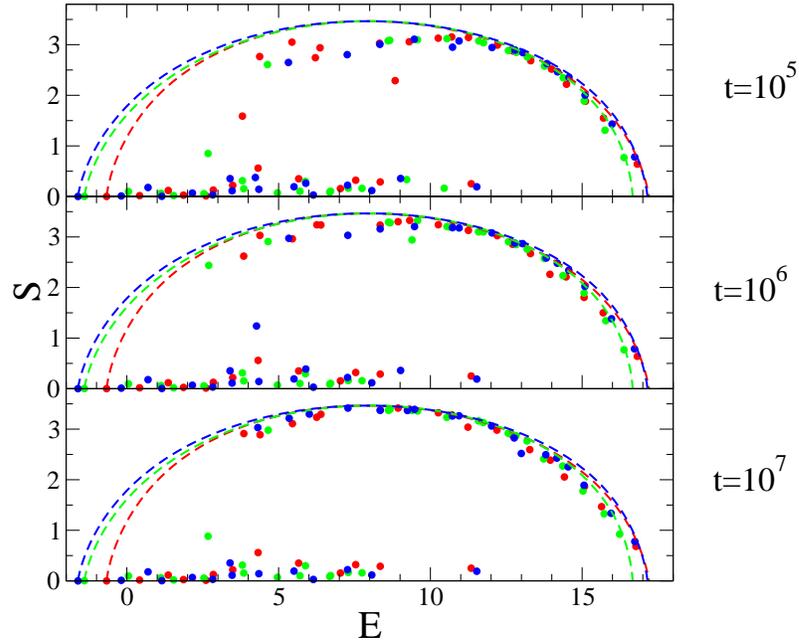}
\end{center}
\vglue -0.2cm
\caption{Dependence of entropy $S$ on energy $E$
at three moments of time $t$ for three disorder
realisations in 1d model $M2$ at
$f=1$, $W=2$, $\beta=0.2$, $N=32$; the averaging is done
in the time interval $\delta t =t/10$. Points show the numerical data
for three disorder realisations (three colors),
dotted curves show the corresponding theoretical
Gibbs distributions.
}
\label{fig4}
\end{indented}
\end{figure}

The dependencies $S(E)$ for 1d lattice models $M1, M2$
are shown in Figs.~\ref{fig3},\ref{fig4}.

For $M1$ we see that at $\beta=2$ the quantum Gibbs works well at
$f=0.5$ and $W=2$. At fixed $W$ an increase of $f$  leads to 
appearance of a significant number of non-thermalized modes
at $f=2$. Indeed, at large $f$ the average distance
 between linear modes
is growing $\Delta \omega \approx f$
and a nonlinear frequency broadening 
$\delta \omega$ becomes to be too small so that the 
nonlinear coupling between
linear modes  starts to be perturbative
and the integrability sets in for larger and larger number of
initially excited modes. 
A similar situation for $3$ oscillators with a
nonlinear coupling had been discussed in \cite{3oscil}.
An increase of disorder from $W=2$ up to $W=4$
reduce the localization length $\ell$
and the number of coupling terms
between linear modes drops.
This leads to a larger number of non-thermalized modes.

For the model $M2$ in Fig.~\ref{fig4} 
we take a relatively small
value of nonlinearity $\beta=0.2$. 
Thus at $m < m_c \approx N/3$ 
we have
a local effective $\beta_{eff} \approx \beta |m - n_0| < 1$,
thus the dynamics remains mainly integrable and
the dynamical thermalization is absent for low energy modes.
However, at $m>m_c \approx N/3$ we have the onset of dynamical
thermalization and the Gibbs law works for high energy modes. 
With the increase of time we see the increase of number of thermalized modes
at $m>m_c$.

\section{Results for 2d lattice models}

\begin{figure} 
\begin{indented}\item[]
\begin{center}
\includegraphics[width=0.80\textwidth]{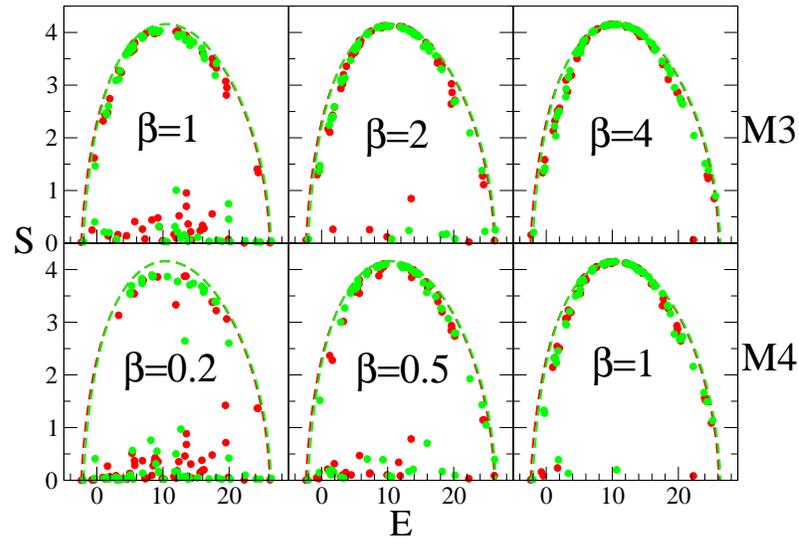}
\end{center}
\vglue -0.2cm
\caption{Dependence of entropy $S$ on energy $E$
at time $t=10^6$ for two disorder
realisations in 2d models $M3$ (top row) 
and $M4$ (bottom row). Here 
$f=1$, $W=2$, the lattice has $8 \times 8$ sites,
the averaging is done
in a time interval $\delta t =t/10$. 
Points show the numerical data, 
dashed curves show the corresponding theoretical
Gibbs distributions, 
the values of $\beta$ are directly given
on the panels.
}
\label{fig5}
\end{indented}
\end{figure}

\begin{figure} 
\begin{indented}\item[]
\begin{center}
\includegraphics[width=0.25\textwidth]{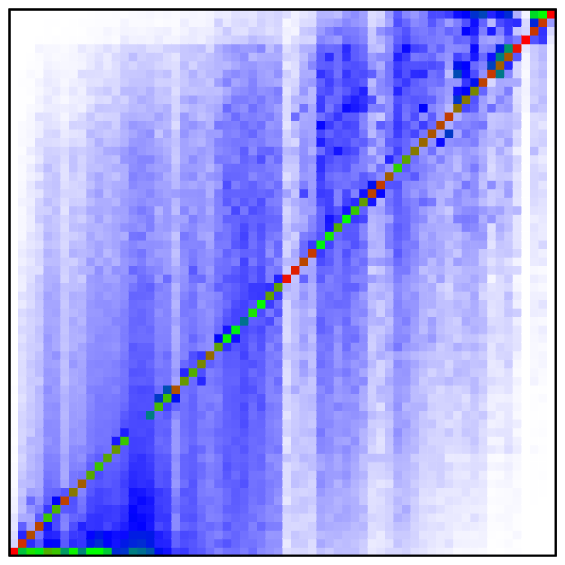}
\includegraphics[width=0.25\textwidth]{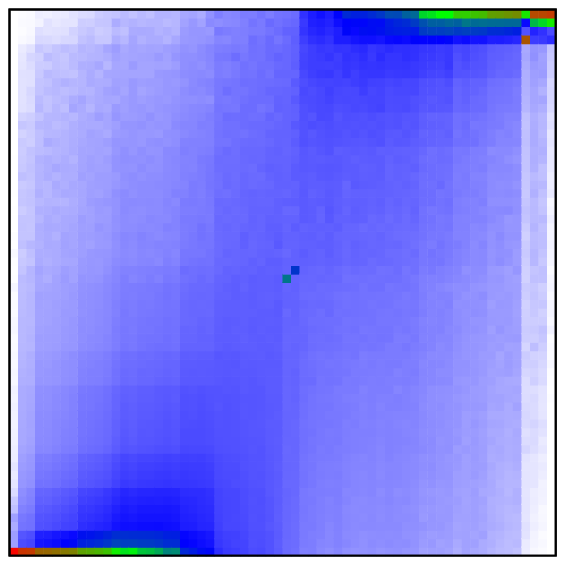}
\includegraphics[width=0.25\textwidth]{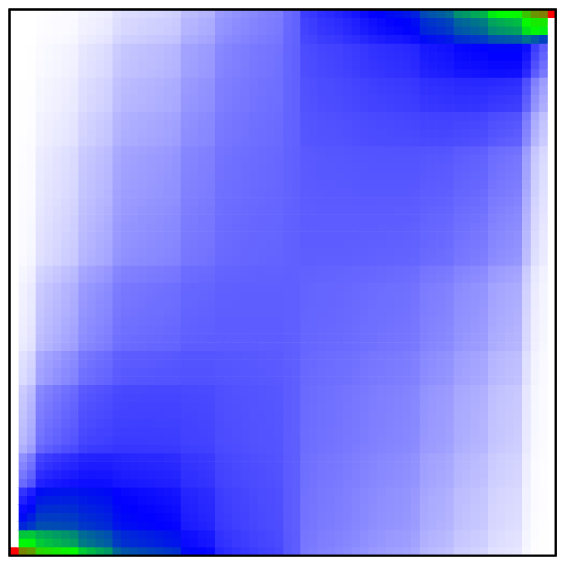}
\end{center}
\vglue -0.2cm
\caption{Time and disorder averaged probabilities $\rho_{m}(m^\prime)$
in mode $m$ for initial state in mode $m^\prime$ for 2d model $M3$ with 
$f=1$, $W=2$, the lattice has $8 \times 8$ sites; 
the probability is proportional to color
changing from maximum (red) to minimum (white).
Panels show the indexes
$1 \leq m' \leq N$ in $x$-axis
and $1 \leq m \leq N$ in $y$-axis. The average is done over
$N_d=10$ disorder realisations 
using the time interval $\delta t=10^6$ starting from time $t=10^6$.  
Left and center panels show the cases with $\beta=1$ and $\beta=4$ 
respectively while right panel 
illustrates theoretical values  
obtained from the quantum  Gibbs distribution (\ref{eq10}). 
}
\label{fig6}
\end{indented}
\end{figure}

The results for 2d lattice models $M3, M4$
are presented in Figs.~\ref{fig5},~\ref{fig6}.
We note that the model $M3$
can be also viewed as a model for 
a nonlinear interaction of laser modes in optical fibers
which $z$-propagation along the fiber is
analogous to the time evolution in our model.
At present the nonlinear dynamics of modes
in laser fiber arrays attracts a significant interest
of optics community (see e.g. \cite{turitsyn1,turitsyn2,turitsyn3}). 

We take here a relative large value of a static field
$f=1$ having in mind to model the 
evolution of the nonlinear Sch\"odinger equation in 
a Sinai billiard. Of course,  the model $M3$
is only an approximation of this physical system.
The obtained results resemble those found for 1d models.
At weak nonlinearity we have a a large fraction of non-thermalized
modes while for $\beta \ge 2$ (in $M3$) and
$\beta \ge 1$ (in $M4$) we find that 
practically all initial conditions with linear eigenmodes
follow the $S(E)$ curve given by the quantum Gibbs anzats.

A more detailed comparison between 
the numerically obtained probabilities $\rho_m$ and the
probabilities given by the Gibbs anzats is shown in Fig.~\ref{fig6}.
For a given disorder realisation we start 
from linear eigenmode $m'$ and numerically determine 
the time averaged probability $\rho_m$ in each
of $N$ linear modes. In addition $\rho_m$ are averaged 
over 10 disorder realisations. The numerical results 
at $\beta=1, 4$, $W=2$, $N=32$ are compares with the theoretical
probabilities of Gibbs anzats obtained for the same
disorder realisations (Fig.~\ref{fig6}). We see that for 
$\beta=1$ (left panel) there is a significant
probability to find non-thermalized modes
well visible as a high density near the diagonal.
However, for $\beta=4$ we have a good agreement with the
probability distribution of Gibbs anzats. 

\section{Results for 1d Klein-Gordon lattice model}

The above lattice models have two exact integrals
of motion being the energy of the system and the total probability.
These models are obtained from a nonlinear Sch\"odinger equation
and hence the appearance of the quantum Gibbs distribution can
be viewed as somewhat natural result 
with the dynamical thermalization
over quantum linear modes produced by a moderate nonlinearity. 
Due to that it is interesting to study the case of KG model which
have only one energy integral.

\begin{figure} 
\begin{indented}\item[]
\begin{center}
\includegraphics[width=0.80\textwidth]{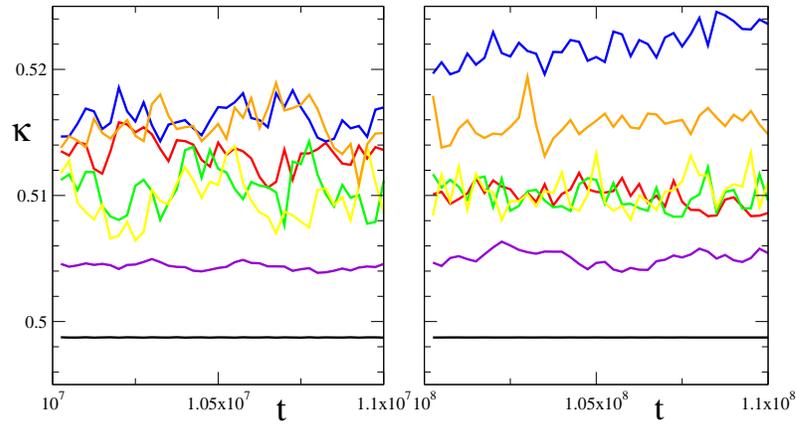}
\end{center}
\vglue -0.2cm
\caption{Time evolution of the norm of $\rho$ averaged in time, 
given by $\kappa=\sum_m \overline{\vert C_m\vert^2}$. 
The averaged time $\delta t$ and starting time $t$ are 
$\delta t=2.5\times 10^5$ and $t\sim 10^7$ for left panel and 
$\delta t=2.5\times 10^6$ and $t\sim 10^8$ for right panel. 
The initial states are taken as normalized eigenstates of the linear 
system with $m=1,6,11,16,21,26,31$
(with corresponding colors: black, red, blue, green, yellow, orange, violet)
 for KG model at 
$N=32$, $W=2$, $f=1/8$ and $\beta=1$.
}
\label{fig7}
\end{indented}
\end{figure} 

To understand the properties of KG model we
note that the eigenmodes of its  linear part
are described by the same linear equations as 
1d Anderson model (see the correspondence description in
\cite{flach2009kg}). To explore this correspondence 
in a deeper way we determine the 
eigenmodes of displacements $u_{lm}$
with eigenfrequencies $\omega_m^2$.
The time evolution of the nonlinear
KG equation (\ref{eq9})
is solved numerically up to times $t=10^8$
for different disorder realizations.
At the initial time $t=0$ we start with
$u_l(t=0)=u_{lm}$ and $p_l(t=0)=0$ ($\sum_l u_{lm}^2=1$).
During the time evolution
we compute the expansion coefficients
$C_m(t)=\sum_l u_l(t) u_{lm}$.
From them we determine the time averaged norm
$\kappa=\sum_{m'} \overline{\vert C_{m'}\vert^2}$
where the averaging is done over
a time interval $\delta t$ around time $t$.
The dependence of $\kappa$ on time
for various initial eigenmodes $m$
is shown in Fig.~\ref{fig7}.
We see that even at very large times $\kappa$
remains approximately constant
with variations remaining on a level of $1-2$ percents.
On average we have $\kappa \approx 1/2$ since a half of
energy is concentrated in the kinetic part
which is taken at zero at $t=0$.
Since $\kappa$ remains an approximate integral of
motion we define the probabilities
$\rho_m =|C_m|^2/\kappa$ so that
their sum is normalized to unity
at a given moment of time
$\sum_m \rho_m=1$. With such a definition of
$\rho_m$ we compute the entropy $S$ 
of the KG model 
via the usual relation (\ref{eq11}).
Of course, this normalization does not affect the
actual values of $u_l(t)$
computed during the time evolution.
The energy $E$ is the 
total energy of (\ref{eq9})
with the initial state being the linear eigenmode
$u_{lm}$ and $p_l=0$. The energies of 
linear modes are $\epsilon_m=\omega_m^2/2$.
With these conditions we can test the validity of 
the Gibbs anzats for the KG model.

\begin{figure} 
\begin{indented}\item[]
\begin{center}
\includegraphics[width=0.80\textwidth]{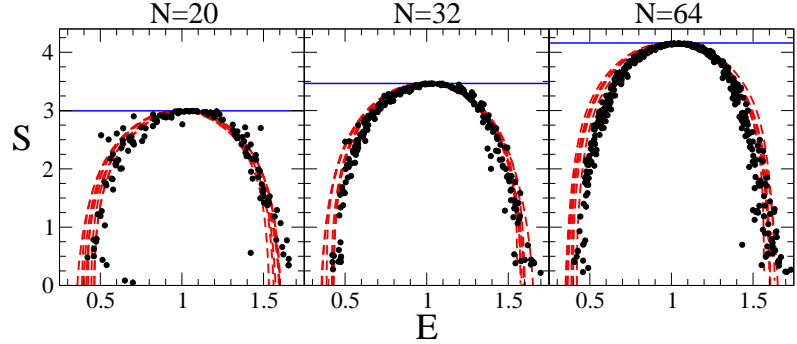}
\end{center}
\vglue -0.2cm
\caption{Dependence of entropy $S$ on energy $E$
at time $t=10^8$ for $7$ disorder
realisations in the KG model at $W=2, \beta=1, f=0$. 
The averaging is done in the time interval 
$\delta t =10^6$. The lattice size is $N=20,32,64$ 
(from left to right respectively).
Points show the numerical data, 
dashed curves show the quantum 
Gibbs distributions for each disorder realization.
Solid blue line 
represents maximum entropy state 
given by a uniform distribution.
}
\label{fig8}
\end{indented}
\end{figure} 

The results for the standard parameters of the KG model
at $\beta=1$, $f=0$, used in \cite{flach2009kg},
are shown in Fig.~\ref{fig8}. At small lattice size $N=20$ the 
fluctuations are present in $S(E)$ dependence but at larger sizes
$N=32, 64$ we find  a good agreement of numerical data with 
the quantum Gibbs anzats.

\begin{figure} 
\begin{indented}\item[]
\begin{center}
\includegraphics[width=0.80\textwidth]{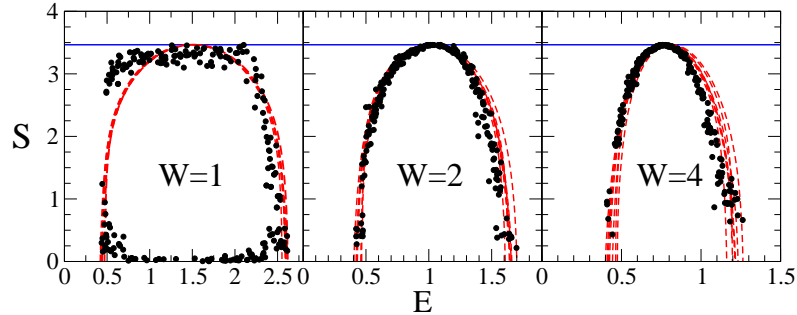}
\end{center}
\vglue -0.2cm
\caption{Same as in Fig.~\ref{fig8}
for $W=1,2,4$ at $\beta=1$, $f=0$, $N=32$;
other parameters are as in Fig.~\ref{fig8}.
}
\label{fig9}
\end{indented}
\end{figure} 

The dependence on effective disorder strength
$W$ is shown in Fig.~\ref{fig9} for the standard
parameters of KG model $\beta=1, f=0$.
We see that at disorder
$W=1$ there is a significant fraction on
non-thermalized states. We attribute this to the
fact that, in 1d Anderson model at such $W$, the localization length
$\ell \approx 96$ becomes much larger than the system
size and the linear modes cross the system practically in a ballistic
way leading to a different onset of chaos.
For $W=2, 4$ we find a good agreement with the Gibbs anzats.

\begin{figure} 
\begin{indented}\item[]
\begin{center}
\includegraphics[width=0.80\textwidth]{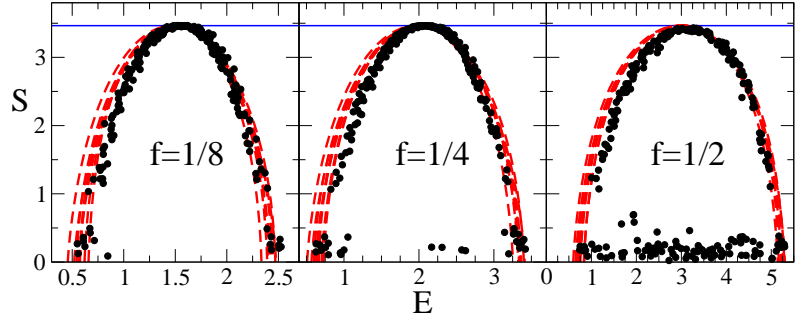}
\end{center}
\vglue -0.2cm
\caption{Same dependence $S(E)$ as in Fig.~\ref{fig9}
for $W=2, \beta=1, N=32$ in KG model and different $f$
shown directly in panels; here $t=10^8$, $\delta t =10^6$,
data are shown for $8$ disorder realisations.
}
\label{fig10}
\end{indented}
\end{figure} 

The dependence of $S(E)$ on $f$ for a fixed $\beta=1$, $W=2$
is shown in Fig.~\ref{fig10}. As for the DANSE type models
discussed above we find that at large $f$ the fraction
of non-thermalized modes becomes significant.
The physical reasons are the same: the average
spacing between linear modes becomes larger than
the nonlinear coupling and the system starts to approach 
an integrable regime.

\begin{figure} 
\begin{indented}\item[]
\begin{center}
\includegraphics[width=0.80\textwidth]{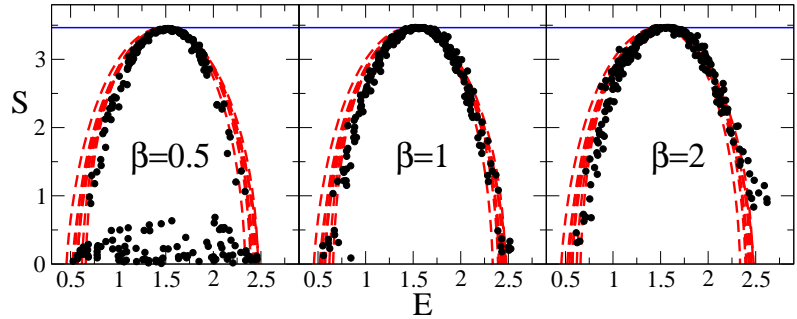}
\end{center}
\vglue -0.2cm
\caption{Same as in Fig.~\ref{fig10}
for $f=0.125, W=2, N=32$ and $\beta$ shown 
directly in the panels; same $8$ disorder realisations.
}
\label{fig11}
\end{indented}
\end{figure} 

The dependence of $S(E)$ on nonlinear parameter
$\beta$ at fixed $f=0.125, W=2$ is shown in Fig.~\ref{fig11}.
At $\beta=0.5$ we still have non-thermalized modes
in the Kolmogorov-Arnold-Moser intergability regime.
The numerical data are in a good agreement 
with the Gibbs anzats at $\beta= 1$
while at $\beta=2$ the deviations become slightly visible.
The deviations become larger for $\beta=4$
(data not shown).
This happens since at large $\beta$ the nonlinear part of Hamiltonian
is not weak or moderate and, hence, it leads to appearance of
significantly nonlinear effects including breathers and
other phenomena. It is possible that the classical energy 
equipartition over linear modes will appear at such larger 
nonlinearities. Thus we find that the quantum Gibbs anzats is
valid inside a certain finite range of nonlinearity 
$\beta_{min} < \beta < \beta_{max}$.

\begin{figure} 
\begin{indented}\item[]
\begin{center}
\includegraphics[width=0.25\textwidth]{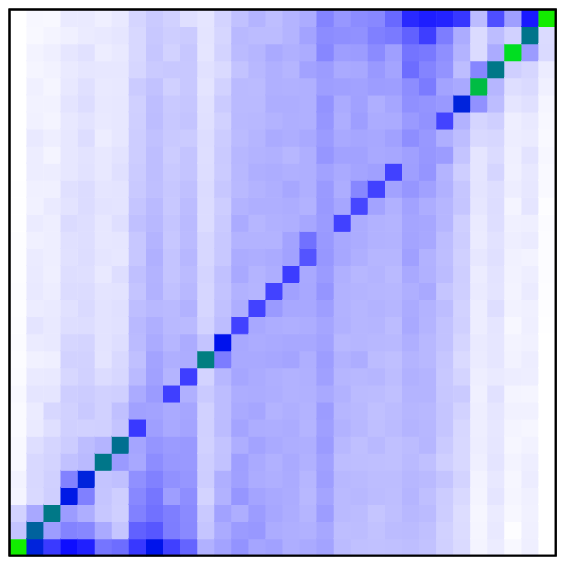}
\includegraphics[width=0.25\textwidth]{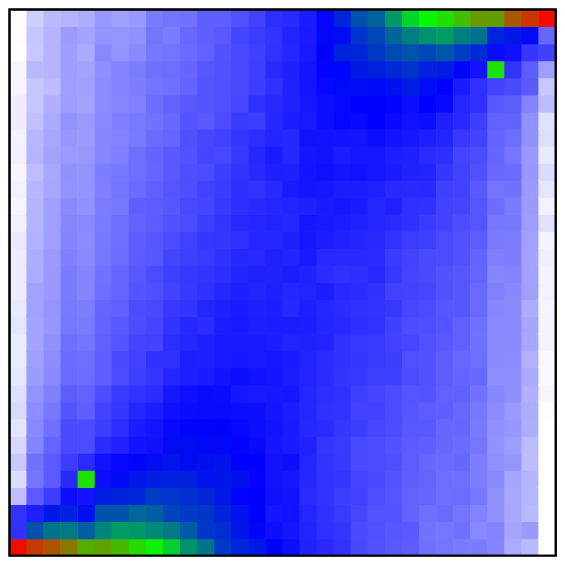}
\includegraphics[width=0.25\textwidth]{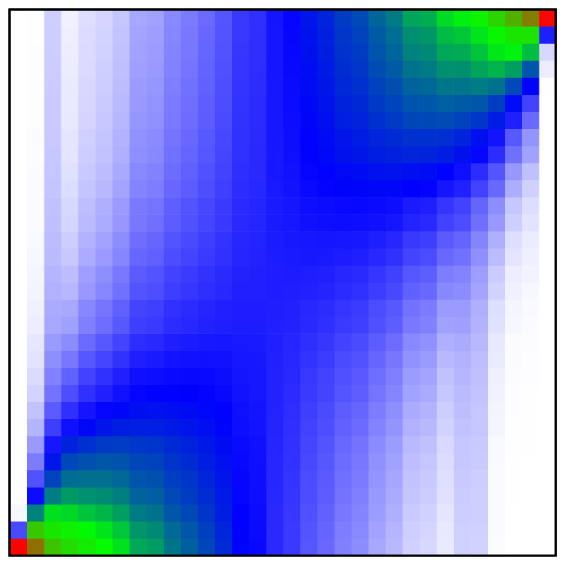}
\end{center}
\vglue -0.2cm
\caption{Same probability distribution $\rho_m(m')$
as in Fig.~\ref{fig6} shown for the KG model
with parameters $f=0.125, W=2, N=32$
at $t=10^8$, $\delta t=10^6$, and $5$ disorder realisations.
Left and center panels show the cases with $\beta=0.5$ and $\beta=1$ 
respectively while right panel 
illustrates theoretical values  
obtained from the quantum  Gibbs distribution (\ref{eq10}). 
}
\label{fig12}
\end{indented}
\end{figure}

At moderate nonlinearities $\beta= 1$ we find
that not only the curve $S(E)$ is well described by
the Gibbs anzats but also the probabilities
$\rho_m(m')$. This fact is illustrated in Fig.~\ref{fig12}
where the probability distributions $\rho_m(m')$,
shown in color, are in a good agreement with 
the quantum distribution of probabilities
given by the theoretical Gibbs distribution.
At small $\beta=0.5$ we have non-thermalized states 
with a higher density at the diagonal similar to Fig.~\ref{fig6}.

\section{Discussion}

In this work we studied numerically the
time evolution in various types of classical lattices with
moderate nonlinearities. We show that at moderate
values of nonlinear parameter $\beta_{min} < \beta < \beta_{max}$
and at large time scales 
the nonlinear interactions  between linear lattice modes
creates a steady state quantum probability distribution over
energies of linear modes. This 
steady state probability distribution
is well described by the quantum Gibbs anzats (\ref{eq10})
being drastically different from the classical steady state
energy equipartition over linear modes expected from
classical thermalization picture.
In a certain sense the nonlinear term generates
a dynamical thermalization in the whole system
with the emergence of the quantum Gibbs distribution.
The appearance of such a quantum statistics
takes place not only in the lattices with
a discrete Sch\"odinger equation (DANSE type),
where energy and norm are both conserved, but
also in other type of lattices which have only
one exact integral of energy.
We argue that in the latter case there is
an approximate conservation of norm
that makes again such nonlinear lattices to
be similar to the DANSE type case.
The emergence of the quantum Gibbs anzats in
nonlinear lattices with only 
one energy integral of motion
allows us to make a conjecture
that the quantum Gibbs anzats is a generic phenomenon
typical for many-mode lattices with weak or moderate nonlinearities.
Indeed, a system of linear oscillators
is effectively equivalent to a certain
effective Schr\"odinger equation
and thus a nonlinear interaction of modes
can drive a generic lattice 
to the quantum Gibbs distribution
via dynamical thermalization.
We think that the further analysis of dynamical
thermalization in nonlinear classical lattices
is of fundamental importance for
a deeper understanding of onset of 
ergodicity and  thermalization in
such systems.

We hope that the phenomenon of dynamical thermalization
described here can be tested in experiments
with cold atoms in optical lattices (e.g. like
in \cite{inguscio,braun}), nonlinear photonic 
lattices (e.g. like in \cite{segev,silberberg})
or optical fiber arrays \cite{turitsyn1,turitsyn2,turitsyn3}
which seems for us to be especially promising.

\ack
We thank A.S.Pikovsky for useful discussions.

\end{document}